\documentclass[twocolumn,showpacs,preprintnumbers,amsmath,amssymb]{revtex4}
\usepackage{epsfig}
\usepackage{dcolumn}
\usepackage{bm}
\DeclareGraphicsExtensions{.eps,.ps,.eps.gz,.ps.gz,.eps.Z,{}}

\newcommand{\dd}{{\rm d}}

\newcommand{\chib}{\overline{\chi}}

\newcommand{\mg}{\langle}
\newcommand{\md}{\rangle}
\newcommand{\phib}{\overline{\phi}}

\newcommand{\mSb}{\overline{{\cal S}}}
\newcommand{\Dirac}{\delta_{\rm Dirac}}
\newcommand{\cc}{\hbox{c.c.}}
\newcommand{\sym}{\hbox{sym.}}
\newcommand{\apjl}{Astrophys. J. Letter}
\newcommand{\physrep}{Physics. Rep.}

\newcommand{\Mpl}{M_{\hbox{Pl.}}}

\def\vk{{\bf k}}
\def\vx{{\bf x}}

\def\dd{{\rm d}}
\def\ii{{\rm i}}

\def\mP{{\cal P}}
\def\mR{{\cal R}}
\def\mS{{\cal S}}

\begin{document}
\title{Non-Gaussianities in extended D-term inflation}

\author{Francis Bernardeau}
\affiliation{Service de Physique Th{\'e}orique,
         CEA/DSM/SPhT, Unit{\'e} de recherche associ{\'e}e au CNRS, CEA/Saclay
         91191 Gif-sur-Yvette c{\'e}dex}
\author{Tristan Brunier}
\affiliation{Service de Physique Th{\'e}orique,
         CEA/DSM/SPhT, Unit{\'e} de recherche associ{\'e}e au CNRS, CEA/Saclay
         91191 Gif-sur-Yvette c{\'e}dex}

\vskip 0.15cm

\begin{abstract}
We explore extensions of hybrid inflationary models in the context
of supersymmetric D-term inflation. We point out that a large
variety of inflationary scenarios can be encountered when the field
content is extended. It is not only possible to get curvaton
type models but also scenarios in which different fields, with
nontrivial statistical properties, contribute to the primordial
curvature fluctuations. We explore more particularly the parameter
space of these multiple field inflationary models. It is shown that
there exists a large domain in which significant primordial
non-Gaussianities can be produced while preserving a scale free
power spectrum for the metric fluctuations. In particular we
explicitly compute the expected bi- and trispectrum for such models
and compared the results to the current and expected observational
constraints. It is shown that it is necessary to use both the bi-
and tri-spectra of CMB anisotropies to efficiently reduce their
parameter space.
\end{abstract}

\pacs{98.80.-k, 98.80.Cq, 98.80.Es,98.80.Hw} \vskip2pc

\maketitle
\section{Introduction}

The new generations of high precision Cosmic Microwave Background
(CMB) measurements opens a new window to the physics of the early
universe and more specifically to the physics of inflation. Up to
now, the shape of the power spectrum has been the prime object of
interest in order to confront theory and observation. Any observed
departure from a scale free, Harrison-Zel'dovich type, power
spectrum would indeed give a precious hint on the shape of the
inflaton potential and subsequently on models of
inflation~\footnote{The recent measurement of
WMAP~\cite{2006astro.ph..3449S} on the primordial index is still a
matter of debate.}. But needless is to say that detection of
primordial non-Gaussianities in the CMB temperature anisotropies
or polarization would also carry precious indications regarding
the inflaton sector.

It has been shown however that in its simplest formulation
inflation does not lead to significant non-gaussianities (NG). This result
has been rigorously established at the level of the bispectrum
over the last few
years~\cite{2003JHEP...05..013M,2005JCAP...06..003S,2004PhR...402..103B}
and recently it has been extended to
trispectrum~\cite{2007JCAP...01....8S,2006PhRvD..74l3519B}. To be
more precise the low level of primordial NG that standard
inflation is expected to induce is unlikely to be detected in
practice due to the subsequent evolution of the metric whether it
is at superhorizon scale, during or after the inflationary phase or
during the complex stages of the
recombination~\cite{2006JCAP...06..024B,2007JCAP...01...19B}.

It appears then useful to have at our disposal sets of models in
which significant primordial non-Gaussianities can be produced,
that is an amount of NG that clearly exceeds what the nonlinear gravitational evolution is
inducing. This appears to be the case for instance for models
built out of actions close to the Dirac-Born-Infeld one
\cite{2004PhRvD..70j3505S,2004PhRvD..70l3505A}. Those models
however rely on non trivial extension of the kinetic part of the
Lagrangian. Although that would be a ground breaking discovery to
see such effects, it remains that the appearance of primordial NG
does not require the use of such non-standard extension of the
low-energy field equation. The curvaton model offers another path
to possible large amount of primordial
NG~\cite{2002PhLB..524....5L}. Here primordial non-Gaussianities
are induced lately, at the time the energy density of the curvaton
field overcome the inflaton decay products. The efficiency of such
a mechanism  therefore depends on sectors of the theory that are
not directly relevant to the inflation.

A third possibility has recently been advocated which is entirely
determined in the inflationary sector. It is provided by some models
of multiple field inflations. The mechanism at play has been
described in various papers
~\cite{2002PhRvD..65j3505B,2002PhRvD..66j3506B}. The difficulty so
far is to identify more precisely models in which such a scenario
could take place. In \cite{2003PhRvD..67l1301B} explicit Lagrangian
(with canonical kinetic terms) were proposed for which significant
non-Gaussianities can be generated during the inflationary phase.
Although it was demonstrated that it was possible to get them during
the whole inflationary stage, the most efficient (and in some sense
most natural) model in which such a possibility was achieved was
based on an hybrid inflation type construction. In this case
significant non-Gaussianities can be transferred to the metric
fluctuations at the very end of inflation, at the time of the
tachyonic instability~\footnote{From this point of view such a model
resembles modulated inflation models, in particular the one
presented in \cite{2004PhRvD..70h3004B}.}. The model proposed in
\cite{2003PhRvD..67l1301B} was however purely phenomenological and
was not motivated by high energy physics constructions.


We propose here to explore a bit further high-energy physics motivations for 
multiple-field hybrid inflation. Hybrid inflationary models
introduced by A. Linde~\cite{1994PhRvD..49..748L} provided a way to
circumvent one fundamental issue in standard inflation with simple
potentials, that is that the Vacuum Expectation Values (VEVs) of the
inflaton field is bound to be of the order of the Planck mass during
the inflationary period~\cite{2000eaa..bookE2135L}. This is the case
in particular with quadratic or quartic potentials. With hybrid
inflation the end of inflation is triggered by a tachyonic
instability and the VEVs of the fields can be made as small as one
wishes (at the expense of a power spectrum index very close to
unity). Hybrid inflation has received some further theoretical
justification because it naturally appears in various high energy
setups. In the context of global supersymmetry, it is indeed
possible to construct potentials which allow a viable inflationary
period, either based on the F- or the D-term of the
potentials~\cite{1994PhRvL..73.1886D,1996PhLB..388..241B}. The aim
of this paper is to take advantage of such theoretical frameworks to
build  models of multiple field inflation and explore their
phenomenological consequences.


The plan of the paper is the following. We first succinctly review
the construction and constraints associated to supersymmetric hybrid
models. We then present possible extensions of these models and
their phenomenological consequences. The amount of non-Gaussianity
they can lead to is precisely computed and confronted to
observational constraints.

\section{Supersymmetric Hybrid inflation}

In a supersymmetric context, two scenarios have been identified
that lead to inflationary models that reproduce hybrid type
potentials. Those scenarios are the F- and D- term inflationary
models~\cite{1999PhR...314....1L}. They allow in particular to have inflation while the field
VEVs remain small compared to the Planck mass. In such a case the
expression of the low energy Lagrangian is safe from corrective
terms originating from supergravity effects (e.g. through the
K\"ahler potential) or high-order curvature terms. This is the
domain we will stick to throughout this paper.

To be more specific, the F- and D-term inflation are based on the
specific global supersymmetry breaking mechanisms. Once symmetry
breaking occurs, the potential has to be corrected for one-loop
radiative correction terms. These corrections are encapsulated in the the formula
of Coleman-Weinberg derived in~\cite{1973PhRvD...7.1888C}. For soft
supersymmetry breaking schemes (where the trace of the mass matrices
remain unchanged)  one is left with logarithmic corrections to the
potential. This is precisely what we have for F- and D-term inflationary models.

The F-term inflation is based on a superpotential of the form~\cite{1994PhRvL..73.1886D},
\begin{equation}
W=\lambda \mS \phib \phi-\mu^2\mS
\end{equation}
where $\mS$, $\phib$ and $\phi$ correspond to the scalar degrees of freedom.
The resulting potential is the following. It comes from F-term contribution only,
\begin{equation}
V=\lambda^2 \vert\mS\vert^2 \left(\vert\phib\vert^2+\vert\phi\vert^2\right)+\vert\lambda\phib\phi-\mu^2\vert^2.
\end{equation}
The energetic landscape it leads to shows that there exists a flat
direction along which the potential is constant (at tree order) but
non-vanishing. It is for a such field configuration that the
inflationary phase can take place. This flat direction is obtained
for $\phi=\phib=0$. It corresponds to a stable minimum (e.g. all
masses are positive) as long as
\begin{equation}
\lambda ^2\vert\mS\vert^2>\mu^2 \lambda,
\end{equation}
that is $\vert\mS\vert > \mS_{c}$ with
$\mS_{c}={\mu}/{\sqrt{\lambda}}.$ Such a vacuum state obviously
breaks global supersymmetry. It implies that corrective terms due to
non-vanishing radiative corrections have to be added to the
potential. At one-loop order, they approximatively read,
\begin{equation}
V_{\rm 1-loop}\approx\frac{ \lambda^4 \mS_{c}^4}{16\pi^2}\log\frac{\vert\mS\vert}{\mS_{c}}.
\end{equation}
This is thanks to this extra contribution that such a setting can
lead to a  viable  inflationary model. It induces a (slow) roll of
the field $\mS$ towards the origin. In this scenario inflation
terminates\footnote{Inflation can actually end earlier} when
$\vert\mS\vert=\mS_{c}$ due to the tachyonic instability that then
appears. During the inflationary phase the potential can be
approximated by,
\begin{equation}
V_{\rm F-term}(\mS)=\lambda^2 \mS_{c}^4\left(1+\frac{ \lambda^2}{16\pi^2}\log\frac{\vert\mS\vert}{\mS_{c}}\right),
\end{equation}
to a very good approximation.

The mechanism that leads to a D-term inflation
 is very similar~\cite{1996PhLB..388..241B}. This model is based on
the introduction of a $U(1)$ symmetry group and a
non-vanishing Fayet-Iliopoulos term leading to a non-zero D-term in the potential. Another family of inflationary
models can then be constructed with the help of three fields, one,
$\mS$, with zero charge, a field  $\phi$ with a positive charge and
a field $\phib$ with a negative one under the $U(1)$ symmetry and
with the following superpotentiel,
\begin{equation}
W=\lambda \mS \phib \phi.\label{SPDterm}
\end{equation}
It leads to the potential,
\begin{eqnarray}
V&=&V_{F}+V_{D}\nonumber\\
&=&\lambda^2 \vert\mS\vert^2 \left(\vert\phib\vert^2+\vert\phi\vert^2\right)+
\lambda^2\vert\phib\phi\vert^2+\frac{g^2}{2}\left(\vert\phi\vert^2-\vert\phib\vert^2+\xi\right)^2.
\end{eqnarray}
In a way similar to the F-term inflation, the potential exhibits a
flat stable direction when,
\begin{equation}
\phi=0,\ \ \phib=0,\ \ \vert\mS\vert>\mS_{c}
\end{equation}
with
\begin{equation}
\mS_{c}=\frac{g}{\lambda}\sqrt{\xi}.
\end{equation}
In this case the inflaton potential is given by,
\begin{equation}
V_{\rm D-term}(\mS)=\frac{\lambda^4 S_{c}^4}{2g^2}\left(1+\frac{g^2}{8\pi^2}\log\frac{\vert\mS\vert}{\mS_{c}}\right).
\end{equation}
This is the context in which we will work in the following.

\subsection{Conditions to have a valid D-term inflation}

A number of conditions should be met for such models to provide us
with valid inflationary models: the number of efolds it leads to
should be large enough and the amplitude of the metric fluctuations
is constrained. These are obviously constraints that are generic for
all inflationary models. There are however two difficulties that one
wants to avoid in the context of D-term inflation. Namely one does
not want the VEVs of the fields to exceed the Planck scale so that
global supersymmetry remains a valid description of the
Lagrangian\footnote{Supergravity corrections can obviously be
included in the expression of the
superpotential~\cite{2006JCAP...07..012J}. The whole validity of a
low energy effective theory approach remains however somewhat
questionable in the absence of a full theory of quantum gravity if
the VEVs of some fields approaches Planck scale.} and the formation
of too massive cosmic strings at the end of the inflationary phase
should also be avoided.

The first requirement leads us to assume that not only $\mS_{c}$ is
small but also that the first term of the inflaton potential, the
constant one, is dominating over the second contribution. This is
possible if
\begin{equation}
g^2\ll1.
\end{equation}
With such a constraint the end of inflation takes place when
$\vert\mS\vert=\mS_{c}$. The number of efolds between horizon
crossing for the modes of interest (the ones that are responsible
for the observed large-scale structure) and the end of inflation can
then be easily computed. It is given by
\begin{equation}
N_{e}=\frac{8\pi^2}{g^2\Mpl^2}(\mS_{*}^2-\mS_{c}^2),
\end{equation}
where $\mS_{*}$ is the modulus of $\mS$ during horizon crossings and $\Mpl$ is the Planck mass, $\Mpl^2=1/(8\pi G)$.
The amplitude of the metric fluctuations $\mR_{*}$ for the modes of
interest is observed to be of the order of $10^{-4}$ (see next section for more details). 
It implies
that
\begin{equation}
\frac{4\pi^2}{\sqrt{3}}\frac{\xi\mS_*}{g\Mpl^3}\approx 10^{-4}.
\end{equation}
These two relations lead to the constraint,
\begin{equation}
\frac{16\pi^4}{3}\frac{\xi^2}{\Mpl^4}\left[ \frac{N_e}{8\pi^2}+\frac{\xi}{\lambda^2\Mpl^2} \right]
\approx 10^{-8}.
\end{equation}
This expression exhibits two regimes. One in which $\xi/(\lambda^2\Mpl^2)\ll
N_{e}/(8\pi^2)$ and then the energy scale of $\xi$ is fixed, of the order of
$10^{-5}\,\Mpl^2$, and one in which $\xi/(\lambda^2\Mpl^2)\gg N_{e}$ and where
$\xi\sim 0.5\,\lambda^{2/3}\,10^{-3}\,\Mpl^2$.  But $\xi$ is precisely the energy
scale of the strings that form at the end of inflation. The
constraint on the string content of the Universe favors the
second case (see \cite{2005PhRvL..94a1303R,2006JCAP...07..012J} for details. It is to be noted also that it might be possible to circumvent this constraint).

It is to be noted that in this latter case $\mS_{*}$ is nearly equal to
$\mS_{c}$. More precisely we have,
\begin{equation}
\mS_{*}\approx\mS_{c}\left(1+\frac{N_{e}g^2\,\Mpl^2}{16\pi^2\,\mS_{c}^2}\right).
 \end{equation}
We will see that this is of importance for the phenomenological consequences of D-term extended models.

\section{Extension of the field content}

Here we now propose to extend the field content involved in the
inflationary sector of the theory. Following a conservative approach
regarding the high energy sector of the theory we allow ourselves to
introduce up to cubic terms only in the superpotential.
For instance a  superpotentiel of the form,
\begin{equation}
W=\lambda_{1} \mS_{1}\phib\phi+\mu_{2}^2\mS_{2},\label{SPCurvaton}
\end{equation}
is a priori legitimate. It involves two light fields that coexist
during the inflationary phase. If $\mu_{2}$ is smaller than the
Hubble value during the inflationary phase then super-Hubble
fluctuations can be generated in the $\mS_{2}$ field. Whether those
fluctuations will be observable or not depends on the subsequent
evolution of the $\mS_{2}$ field. In particular, because $\mS_{2}$
is a massive field, its energy density can eventually dominate over
the inflaton decay products if those are relativistic particles. The
$\mS_{2}$ field can then imprint its fluctuations on the metric
fluctuations before it finally decays. This is the curvaton
mechanism. This possibility, and its phenomenological consequences,
have been extensively described in the
literature~\cite{2002PhLB..524....5L}. They depend largely on the
coupling of the extra fields to the other fields of the theory. Such
a theory is then not entirely predictive from the sole knowledge of
the inflationary sector of the theory.

\subsection{Multiple field models}

The expression (\ref{SPCurvaton}) is however not the only possible
extension  of (\ref{SPDterm}). For instance nothing prevents the
introduction of  multiple light fields that are coupled together to
the same charged $U(1)$ fields,
\begin{equation}
W=\sum_{i}\frac{\nu_{i}}{3}\mS_{i}^3+\lambda\left(\sum_{i}\alpha_{i}\mS_{i}\right)\phib\phi.
\end{equation}
Obviously if $\nu_i$ is small enough, the corresponding $\mS_i$ field can participate in the 
inflaton (depending on its initial VEV). The corresponding upper bound for $\nu_i$  for such a possibility to occur is defined in such a way that, when, say, the VEV of $\mS_i$ is below the Planck scale, 
the contribution of the quartic potential it induces is
negligible against the radiative correction term. It leads to the
constraint,
\begin{equation}
\nu^2_{i}\ll \lambda^4.
\end{equation}
The fields for which $\nu_{i}$ is above this bound will rapidly roll towards the origin
but they still can 
develop  significant super-Hubble fluctuations as long as $\nu_{i}$
is smaller than unity. This would not be the case otherwise.

It is then possible to distinguish three sets
of fields, those with a very small coupling constant $\nu$; they
can potentially be part of the inflaton field; the ones with
intermediate values ; they will not contribute to the inflaton but
still develop significant super-Hubble fluctuations and finally
those with a large coupling constant will not develop any
significant fluctuations and will not play any role. In the
following we will be interested in  the second set of
fields.

To simplify the presentation let us assume that there is one field
$\mS_{1}$ with  a vanishing coupling constant $\nu_{1}$ and one
field $\mS_{2}$ with a large - but still smaller than unity -
$\nu_{2}$.
\begin{widetext} 
Then the potential takes the form,
\begin{eqnarray}
V&=&V_{\hbox{1-loop}}
+\lambda^2\left\vert \cos\theta\mS_{1}+\sin\theta\mS_{2}\right\vert^2\left(\vert\phib\vert^2 + \vert\phi\vert^2\right)
+\lambda^2\cos^2\theta\,\vert\phi\vert^2\vert\phib\vert^2
+\vert\nu_2\mS_2^2+\lambda\sin\theta\,\phi \phib\vert^2
\nonumber\\&&
+
\frac{g^2}{2}\left(\vert\phi\vert^2-\vert\phib\vert^2+\xi\right)^2
\end{eqnarray}
where $\theta$ is the mixing angle of $\mS_{1}$ and $\mS_{2}$
encoded in the  $\alpha_{i}$ parameters.
\end{widetext}
As for the previous case, $\phi$ is a very massive field whose VEV is driven to 0. The previous expression
then simplifies and we are left with the effective potential,
\begin{eqnarray}
V&=&V_{\hbox{1-loop}}+\nu_2^2\vert\mS_2\vert^4
+\lambda^2\left\vert \cos\theta\mS_{1}+\sin\theta\mS_{2}\right\vert^2\vert\phib\vert^2
\nonumber\\
&&+
\frac{g^2}{2}\left(-\vert\phib\vert^2+\xi\right)^2
\end{eqnarray}
involving the (complex scalar) fields $\mS_1$, $\mS_2$ and $\phib$.  We see that it involves a new term, the self coupling term of the $\mS_2$ field. It drives the VEV of this field to 0 during the inflationary period. The expression of $V_{\hbox{1-loop}}$ is then left unchanged compared to the single inflationary field case. As for the curvaton model, the $\mS_2$ direction behaves like an isocurvature direction during the inflationary period. Because of the mixing term in the $\mS_i$-$\phib$ coupling though, the $\mS_2$ fluctuations can lead to  visible effects irrespectively of the subsequent evolution of this field.
To be more precise, the inflation terminates when
$\cos\theta\mS_{1}+\sin\theta\mS_{2}$ reaches the critical value
$\mS_{c}$. In the context we are considering, the inflationary
period then ends almost instantaneously because of the tachyonic
instability that subsequently develops
\cite{2001PhRvL..87a1601F,2001PhRvD..64l3517F}.
The end time of inflation is then modulated by 
both fluctuations in the $\mS_{1}$ and $\mS_{2}$ directions. This is the mechanism with which 
initial isocurvature fluctuations are transferred into the adiabatic modes.

The metric fluctuations that are then induced are a combination  of
$\mS_{1}$ and $\mS_{2}$ fluctuations. As long as the only leading
order metric effects are taken into account, the formal expression
of the metric fluctuations can be easily computed using the $\delta
N$
formalism~\cite{1985JETPL..42..152S,1996PThPh..95...71S,1998PThPh..99..763S}.
The induced metric fluctuations then read
\begin{widetext}
\begin{eqnarray}
\delta N(\delta\mS_{1},\delta\mS_{2})&=&\int_{\hbox{inflation traj.}} H(\mS^{(0)}_{1}+
\delta\mS_{1},\delta\mS_{2})\dd t
-\int_{\hbox{inflation traj.}} H(\mS^{(0)}_{1},0)\dd t,
\end{eqnarray}
where $\mS^{(0)}_{1}$ is the zero mode trajectory of the field.
In the context we are interested in the expression of $\delta N$ can easily be explicited
at leading order in the field fluctuations.
It is given by,
\begin{eqnarray}
\delta N&=&-\left.\frac{3H^2}{V_{,\varphi}}\right\vert_{\hbox{Horizon crossing}}\delta\varphi
+\left.\frac{3 H^2}{V_{,\varphi}}\right\vert_{\hbox{end of inflation}}\tan\theta\,\delta\chi_1.\label{dNexp}
\end{eqnarray}
\end{widetext}
where $\varphi$ and $\chi$ are the (canonically defined)
fluctuations  of the fields $\mS_{1}$ and $\mS_{2}$ in the direction
(in the complex plane) of $\mS^{(0)}$. This direction, without loss
of generality, can be defined as the real axis,
\begin{equation}
\varphi=\sqrt{2}\ \hbox{Re}(\mS_{1}),
\end{equation}
and  the real part of $\mS_{2}$,
\begin{equation}
\chi_{1}=\sqrt{2}\ \hbox{Re}(\mS_{2}),\ \ \hbox{and}\ \ \chi_{2}=\sqrt{2}\ \hbox{Im}(\mS_{2}).
\end{equation}
The imaginary parts of those fields are genuine degree of freedom
that  will develop super-Hubble correlations as well. They will
affect though the expression of $\delta N$ at quadratic order only,
as any other couplings to the metric would. At the level of our
description the imaginary part of $\varphi$ will not play any role.
The one of $\chi$ will do however, because it affects the
non-Gaussian properties of $\chi_{1}$ as we will discover.

What is important here to realize is that in the parameter domain
favored by the constraints on the cosmic strings contribution to the CMB anisotropies as discussed previously, the two
coefficients, $\left.{3H^2}/{V_{,\varphi}}\right\vert_{\hbox{horizon
crossing}}$ and $\left.{3 H^2}/{V_{,\varphi}}\right\vert_{\hbox{end
of inflation}}$ are, almost, equal. Moreover  the two fields are independent
of one another; their fluctuations have roughly the same spectrum,
to the difference in their mass, e.g. $\eta$ parameter. The
generated metric fluctuations are then the superposition  of these
two contributions the relative weight is driven by a free parameter,
the mixing angle between the two fields, $\theta$. The amplitude of
the induced metric fluctuations per unit log scale in $k$,
$\mP_{0}$, is then typically given by~\footnote{The amplitude of the
power spectrum of the field fluctuations is $H^2/(2k^3)$.},
\begin{equation}
\mP_{0}\sim \left(1+\tan^2\theta\right)^{1/2}
\frac{3H^3}{2\,V_{,\varphi}}=
\frac{3H^3}{2\cos\theta\ V_{,\varphi}}.
\end{equation}
This is this number that is constrained by the observations:
$\mP_{0}$  is of the order of $2\times10^{-4}$ (see for instance
\cite{2000cils.book.....L}.)

One can be a bit more precise. The first term is expected to be a random field of power spectrum
of index,
\begin{equation}
n_{1}=1-6\epsilon+2\eta,
\end{equation}
and the second of power spectrum index,
\begin{equation}
n_{2}=1-2\epsilon+2\eta_{1},
\end{equation}
where $\epsilon$ is the same in the two cases (it is due to the
variation  of $H$),
\begin{equation}
\epsilon=-\frac{\dot H}{H^2},
\end{equation}
  and $\eta$ are the masses of the fields in
units of $H$ in respectively the adiabatic direction and the
transverse direction,
\begin{eqnarray}
\eta&=&\frac{\Mpl^2}{V}\ \frac{\partial^2 V}{\partial \varphi^2}\\
\eta_1&=&\frac{\Mpl^2}{V}\ \frac{\partial^2 V}{\partial \chi_1^2}.
\end{eqnarray}

The resulting index, in first order in
slow-roll parameter is then,
\begin{equation}
n=1+\cos^2\theta(2\eta_{1}-6\epsilon)+\sin^2\theta(2\eta_{2}-2\epsilon).
\end{equation}
These parameters however vanish as soon as $g$ is small and we are left with a scale free Harrison-Zel'dovich spectrum in all cases.

\subsection{The induced non-Gaussianities}

The most interesting consequences of such a family of models comes
from the  fact that they can potentially induce significant
non-Gaussianities. By significant we mean that such
non-Gaussianities can be much larger than what generic single field
inflation predicts. The amount of non-Gaussianities  which is
generated is actually a direct consequence of the formula
(\ref{dNexp}). Indeed, the self couplings of the field $\chi$ can be
large. We are typically in a situation described in
\cite{2002PhRvD..66j3506B}. Those intrinsic non-Gaussianities are
not due to any coupling of the field to the metric but to the
self-coupling this scalar degree of freedom in a quasi-de Sitter
background.

\begin{figure}
\centerline{
\epsfig {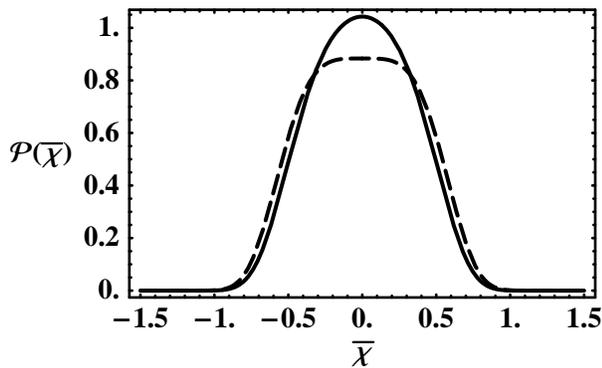}}
\caption{Shape
of the one-point distribution function of $\chib_{1}$  for $H=1$ and
$\nu=1$. The solid line corresponds to the equation
(\ref{PDFchib1exp}) where the imaginary part of $\chib$ has been
integrated out. The dashed line corresponds to the case where
$\chib$ is real.} \label{PDFchib1}
\end{figure}

In particular, it is clear that $\delta \mS_{2}$ can develop
significant  high order correlation function, its four point can be
computed explicitly. Finite volume effects though can induce as well
non-vanishing three-point function. As shown in
\cite{2002PhRvD..66j3506B} it amounts to shift the minimum of the
field to a non zero value. The effective potential in the transverse
directions is actually expected to be
\begin{equation}
V(\chi_{1},\chi_{2})=\frac{\nu^2}{4}\,\left[(\chi_{1}+\chib_{1})^2+
(\chi_{2}+\chib_{2})^2\right]^2,\label{newpot}
\end{equation}
where $\chib_{1}$ and $\chib_{2}$ are in essence random variables
but that  can be considered as fixed for our observable universe. A
classical stochastic approach can be applied to infer the
probability distribution function of its value. The Fokker equation
for the joint distribution of $\chib_{1}$ and $\chib_{2}$ can be
derived from the joint evolution of $\chib_{1}$ and $\chib_{2}$. It
is given by,
\begin{widetext}
\begin{eqnarray}
\frac{\partial \mP}{\partial t}&=&\frac{H^3}{8\pi^2}
\left(\frac{\partial^2 \mP}{\partial \chib_{1}^2}+\frac{\partial^2 \mP}{\partial \chib_{2}^2}
\right)+\frac{1}{3H}\left[\frac{\partial}{\partial \chib_{1}}
\left(\frac{\partial V(\chib_{1},\chib_{2})}{\partial \chib_{1}}\mP\right)
\right.
\left.+\frac{\partial}{\partial \chib_{2}}
\left(\frac{\partial V(\chib_{1},\chib_{2})}{\partial \chib_{2}}\mP\right)
\right],
\end{eqnarray}
\end{widetext}
where $t$ is the physical time.
The late time solution of this equation is
\begin{equation}
\mP(\chib_{1},\chib_{2})=
\frac{\sqrt{2} \nu}{H^2 \sqrt{3\pi }}
\exp\left[-\frac{2\pi^2\nu^2\left(\chib_{1}^2+\chib_{2}^2\right)^2}{3 H^4}\right],
\end{equation}
for the correctly normalized distribution function. For the
bispectrum, we  will see that the parameter of interest is
$\chib_{1}$. Its distribution function, once marginalized over
$\chib_{2}$ is given by,
\begin{equation}
\mP(\chib_{1})=\frac{2 \nu \vert\bar{\chi }_1\vert
   }{H^2 \sqrt{3 \pi }}\,K_{\frac{1}{4}}
   \left(\frac{\pi ^2 \nu ^2 \bar{\chi }_1^4}{3 H^4}\right)\,
   \exp\left[-\frac{\pi ^2 \nu ^2 \bar{\chi }_1^4}{3 H^4}\right] ,
   \label{PDFchib1exp}
\end{equation}
where $K_{\frac{1}{4}}$ is the modified Bessel function of the
second kind  of index $1/4$. The shape of this PDF is shown on Fig.
\ref{PDFchib1} for $H=1$ and $\nu=1$.

For generic initial conditions, and for a large enough number of
efolds prior  to the horizon crossing of the modes of interests,
$\chib_{1}$ is expected to be distributed according to
(\ref{PDFchib1exp}). The excursion domain for $\chib_{1}$ is
therefore typically $[-0.6 H/\sqrt{\nu},0.6 H/\sqrt{\nu}]$. As
suggested before $\chib_1$ can be seen as a free variable in a
situation similar to that encountered in the curvaton scenario. Note
that, from the potential (\ref{newpot}), the field $\chi_{1}$
acquires  a mass equal to $6^{1/2} \nu \chib_{1}$. It is generically
small compared to $H$ ensuring that the field can indeed develop
super-Hubble fluctuations.

\subsection{The perturbative regime}

\begin{figure*}
\centerline{ \epsfig {figure=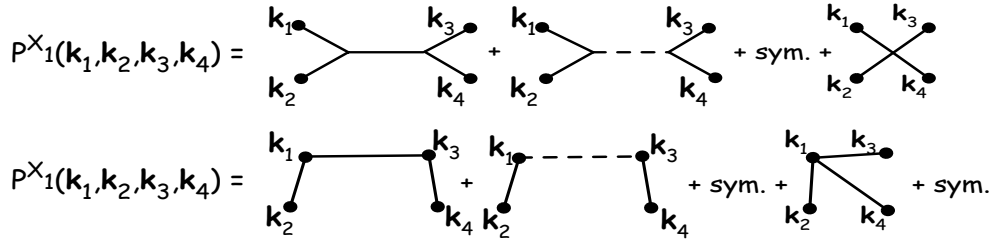,width=14cm}}
\caption{Diagrammatic representation of the contributions to the
connected  four point functions of the $\chi_{1}$ component of the
complex field $\chi$. The upper line show the Feynman type diagrams
that have to be considered in a quantum field approach (for details
see for instance \cite{2005PhRvD..72d3514W}). For tree order
calculations one can equivalently use a classical approach (provided
the initial stochastic fields have the properties derived from the
free field quantum solutions).  The bottom line shows the resulting
diagrams where each vertex point then represents a given order in
the initial field. The mode dependence of those terms can then
easily be derived in the super-horizon limit: the three diagrams
that are represented give respectively $P^{\chi_{1}}(k_{2})
P^{\chi_{1}}(\vert \vk_{1}+\vk_{2}\vert)P^{\chi_{1}}(k_{4})$,
$P^{\chi_{1}}(k_{2}) P^{\chi_{2}}(\vert
\vk_{1}+\vk_{2}\vert)P^{\chi_{1}}(k_{4})$, $P^{\chi_{1}}(k_{2})
P^{\chi_{1}}(k_{3})P^{\chi_{1}}(k_{4})$.} \label{diags}
\end{figure*}

We are interested here in the high order (e.g. three- and
four-point)  correlation functions of the Fourier modes of the field
$\chi$ and consequently $\delta N$. For a field $\sigma(\vx)$ we
define its Fourier transform $\sigma(\vk)$ by,
\begin{equation}
\sigma(\vk)=\int\frac{\dd^3\vx}{(2\pi)^3}\sigma(\vx)\,e^{-\ii\vk\cdot\vx}.
\end{equation}
The power spectrum $P^{\sigma}(k)$ and the higher order spectrum
$P_{n}^{\sigma}(k)$ are defined by
\begin{eqnarray}
\mg\sigma(\vk_{1})\sigma(\vk_{2})\md&=&\Dirac(\vk_{1}+\vk_{2})P^{\sigma}(\vk_{1}),\\
\mg\sigma(\vk_{1})\dots\sigma(\vk_{n})\md_{c}&=&\Dirac(\vk_{1}+\dots+\vk_{n})P_{n}^{\sigma}(\vk_{1},\dots,\vk_{n}),
\end{eqnarray}
where $_{c}$ denotes the connected parts.

A full quantum calculation can be done at tree order in the
perturbative  regime (see last section for details). In a de Sitter background the exact shape can
be obtained for the three and connected part of the four-point
function for the potential (\ref{newpot}). In the super-Hubble
regime they read~\footnote{The original calculation is to be found
in \cite{1993ApJ...403L...1F}.},
\begin{equation}
P_{3}^{\chi_{1}}(\vk_{1},\vk_{2},\vk_{3})=-
\frac{2\,\nu^2 N_{e}\chib_{1}}{H^2}\left[P^{\chi_{1}}(k_{1})P^{\chi_{1}}(k_{2})+{\rm
sym.}\right]\label{bispecmS}
\end{equation}
where $N_{e}$ is the number of efolds between horizon crossing and
the end  of inflation.

As stressed in previous papers, what drives the amplitude of
fluctuations  is the value of $\nu^2 N_{e}$ if $\mSb$ is of the
order of $H$. This result can be expressed phenomenologically
through an $f_{\rm NL}$ parameter. It is defined in such a way that
the bispectrum $P^{\delta N}_{3}(\vk_{1},\vk_{2},\vk_{3})$ of the
metric fluctuation reads,
\begin{equation}
P^{\delta N}_{3}(\vk_{1},\vk_{2},\vk_{3})=2f_{\rm NL}\left[P(k_{1})P(k_{2})+{\rm
sym.}\right],
\end{equation}
when written in terms of the metric power spectrum.
Relation (\ref{bispecmS}) implies that,
\begin{equation}
f_{\rm NL}=-\nu^2\,N_{e}\,\frac{\chib_{1}}{H}\,\frac{\sin^3\theta}{2\mP_{0}}.
\end{equation}
From the adopted definition of $f_{\rm NL}$, the metric fluctuation
will be  significantly non-Gaussian if $\mP_{0}\,f_{\rm NL}$
approaches unity whereas standard inflationary physics implies that
$f_{\rm NL}$ is of the order of unity\,\cite{2004PhR...402..103B}.

Within the framework of this model, it is also possible to compute
any higher  order correlation functions. When $\chib_{1}$ and
$\chib_{2}$ are both taken into account, there are two terms
contributing (and from a field theory point of view, it means that
there are three- as well as four-leg vertices that contribute).

\begin{widetext}
They read,
\begin{equation}
P_{4}^{\chi_{1}}(\vk_{1},\dots,\vk_{4})=4\nu^4\,N_{e}^2\,\frac{9\chib_{1}^2+\chib_{2}^2}{9H^4}
\left[P(k_{1})P(\vert\vk_{1}+\vk_{2}\vert)P(k_{3})+{\rm
sym.}\right]-2\nu^2\,N_{e}\,\frac{1}{H^2}\left[P^{\chi_{1}}(k_{1})P^{\chi_{1}}(k_{2})P^{\chi_{1}}(k_{3})+{\rm
sym.}\right].
\end{equation}
This leads to
\begin{equation}
P_{4}^{\delta N}(\vk_{1},\dots,\vk_{4}) =4f_{\rm NL}^2\left(\frac{9\chib_{1}^2+\chib_{2}^2}{9\chib_{1}^2\sin^2\theta}\right)\left[P(k_{1})P(\vert\vk_{1}+\vk_{2}\vert)P(k_{3})+{\rm
sym.}\right]+6g_{\rm NL}\left[P(k_{1})P(k_{2})P(k_{3})+{\rm
sym.}\right].
\end{equation}
\end{widetext}
The two terms of this equation correspond to two different
geometries. The  coefficient $g_{\rm NL}$ reads,
\begin{equation}
g_{\rm NL}=-\nu^2\,N_{e}\,\frac{\sin^4\theta}{12\mP_{0}^2}.
\end{equation}

Note that the expression of the bispectrum and the second term of
the  tri-spectrum is what a nonlinear transform of the type,
\begin{equation}
\delta N=\delta N+f_{\rm NL}\left(\delta N\right)^2+g_{\rm NL}\left(\delta N\right)^3+\dots
\end{equation}
would have given. Such a transform would however lead to $4f_{\rm
NL}^2$  for the coefficient of the first term of the tri-spectrum.
The difference is due to the fact that the resulting metric
fluctuations are actually the superposition of different fields, not
only the nonlinear transform of a single one.

\begin{figure*}
\centerline{ \epsfig {figure=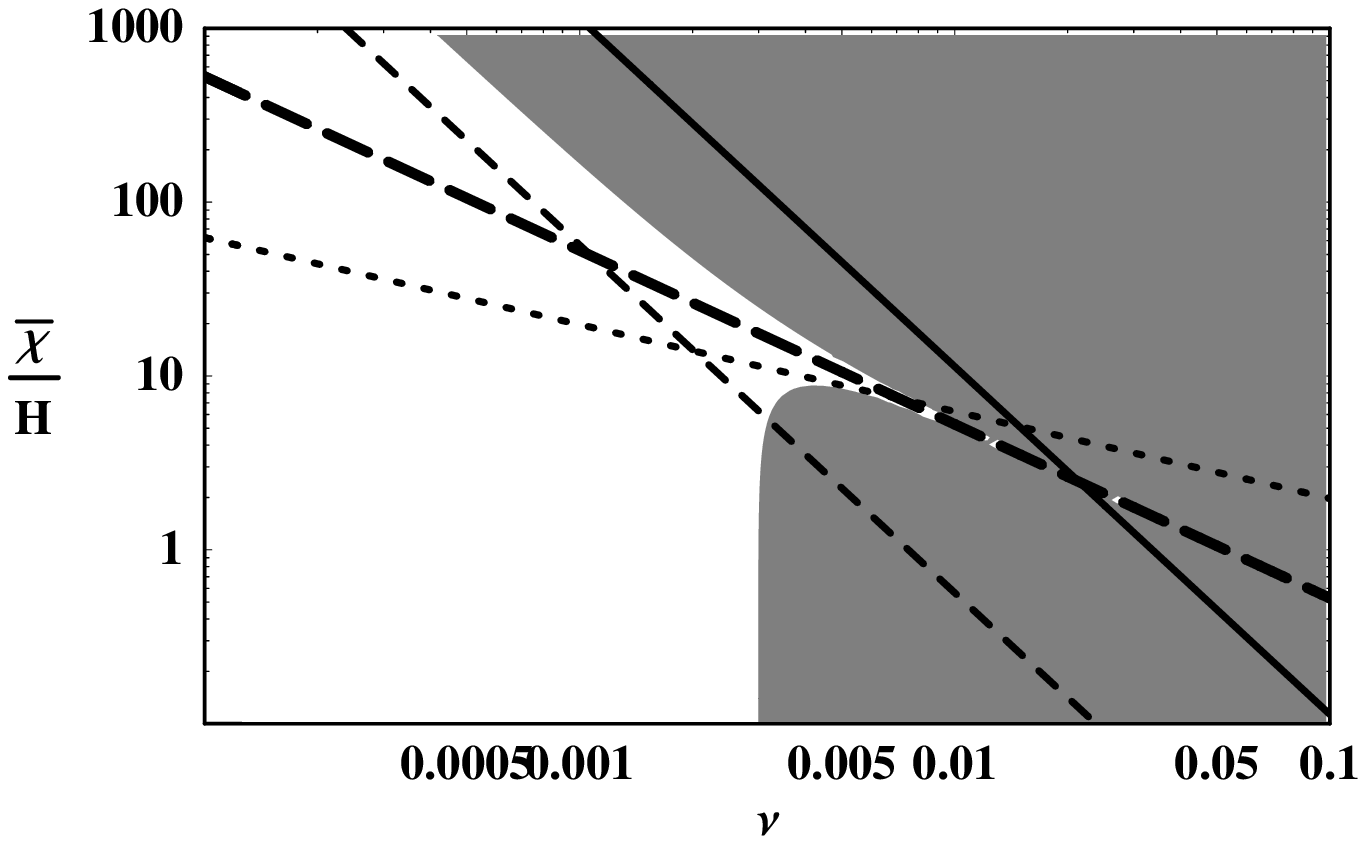,width=8cm}\epsfig {figure=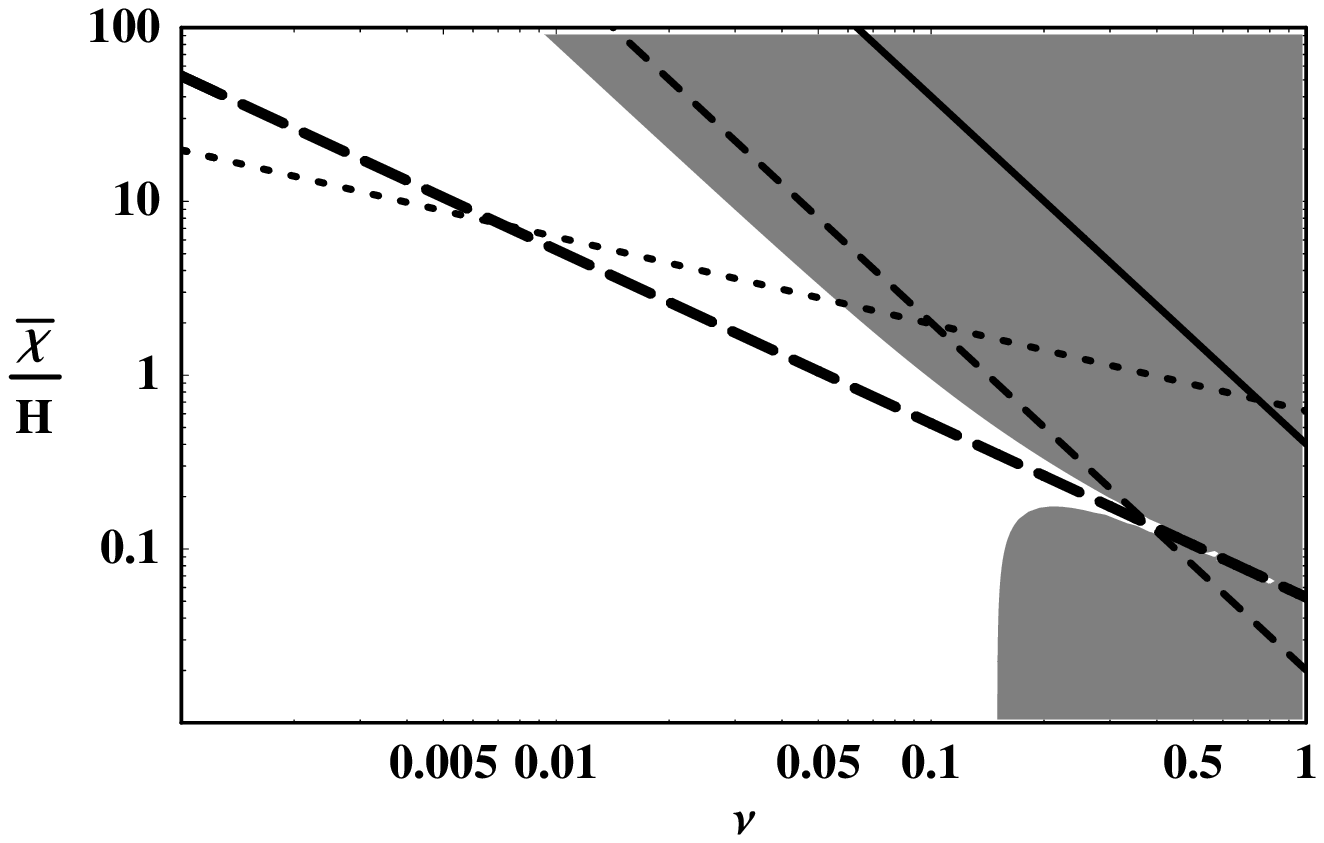,width=8cm}}
\caption{Exclusion diagrams for parameters $\nu$ and $\chib$ for $\theta=\pi/4$ (left panel) and for $\theta=0.1$ (right panel). The locations of the  dotted lines  where $\chib$ is equal to its expected one $\sigma$ fluctuation. The gray areas and solid or short dashed lines correspond to the exclusion zones,  obtained by WMAP (solid line) or expected by Planck (short dashed for bispectrum, gray areas for tri-spectrum). The bispectrum constraint corresponds to a straight line (of slope $-2$); the trispectrum is more complicated due to two competing terms in the trispectrum. The long dashed is the location where the terms cancel. We adopted the results of \cite{2006PhRvD..73h3007K} on the upperbounds the Planck mission is expected to provide, $f_{\rm NL}=5$ and $\tau_{\rm NL}=560$.}
\label{ExclusionMap}
\end{figure*}

The relative weight of these two contributions depends in particular
on  the value of $\mSb$, and therefore on the peculiar realization
of our own universe. These results are valid only in the
perturbative regime, e.g. with respect to the coupling constant. It
appears that the expansion parameter is different during sub-Hubble
physics and during super-Hubble evolution. In the former case,
$\nu^2$ is the parameter to be small. When it is large the field
fluctuations simply vanish away. At super-Hubble scale however one
finds that the effective expansion parameter is actually $\nu^2
N_{e}$ which is obviously much larger than $\nu^2$. It limits the
formal validity of the previous expressions. It also opens a new
specific phenomenological domain the tentative description of which
is the subject of the next section.

In Fig. \ref{ExclusionMap}, we compare the computed amplitude of
the bi- and tri-spectrum to the current (with WMAP) and expected
(with Planck) constraints. The left panel corresponds to a mixing
angle $\theta=\pi/4$, that corresponds to an equal contribution of
the two fields to the metric fluctuations, and the right panel to a
small mixing angle, $\theta=0.1$. We restricted the parameter space
to two variables, $\nu$ and $\chib_{1}$, assuming $\chib_{2}=0$ (it
anyway contributed only weakly to the amplitude of the
tri-spectrum). The dotted lines show the location of the expected
generic value for $\chib_{1}$. Values that would differ too much (in
logarithmic space) from this location would demand some fine tuning
in the peculiar random realization of the universe we live in.

In both cases the thick solid line shows the current constraint
provided by the amplitude of the bispectrum. It naturally limits a
combination of $\chib_{1}$ and $\nu$. When the mixing angle is
large, left panel, it generically leads to small values of $\nu$.
The use of a perturbation theory is then fully justified. When the
mixing angle is small the observations are much mess efficient in
constraining $\nu$. It can be as large as unity. It is then
necessary to reconsider the calculations that have been done to take
into account non-perturbative aspects of the super-Hubble evolution
of the fields.

The Planck mission is expected to provide us with much more
stringent  constraints not only on the bi-spectrum but also on the
amplitude of the tri-spectrum. In Fig. \ref{ExclusionMap} the dashed
lines show the location of the bispectrum constraint provided by
Planck. Regarding the tri-spectrum  we have not attempted to take
into account the different geometrical patterns that appear in its
theoretical expression. As a result the amplitude of the trispectrum
results of the summation of two terms of opposite signs. There is
therefore a location in the parameter space where the tri-spectrum
effectively vanishes. It is shown as a long dashed line. The
constraint that observations would provide is given in terms of
$\tau_{\rm NL}$ which is set to be equal to $6g_{\rm NL}+4f_{\rm
NL}^2/\sin^2(\theta)$. The gray area is the region that Planck could
be able to exclude according to \cite{2006PhRvD..73h3007K}. It
appears clearly that for low values of $\chib_{1}$, the tri-spectrum
is more effectient in constraining the amplitude of the coupling
constant $\nu$. In the rare event tails for $\chib_{1}$, and along
the cancelation line, the bispectrum is more efficient. The two
observations appear therefore very complementary. For low values of
$\theta$ though, it is obviously more difficult to get strong
constraints of the coupling constant $\nu$. In this case the field
$\mS_{2}$ could actually induce primordial NG when it enters a
classical non-perturbative regime.

\subsection{The non-perturbative regime}

\begin{figure}
\centerline{ \epsfig {figure=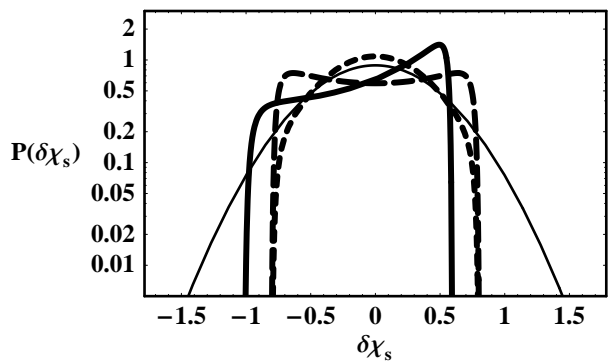,width=8cm}}
\caption{Example of shapes of the one-point PDF of the local value of $\chi$ in case it underwent a non-perturbative evolution at super-horizon scales. The plots correspond to parameter values $\nu N_{e}^{1/2}=1.5$; the long dashed line to  $\chib=0$; the solid line to $\chib_{1}=0.5$, $\chib_{2}=0$ and the short dashed line to $\chib_{1}=0$, $\chib_{2}=2$ (values of $\chib$ are given in units of $H$). The thin solid line is a Gaussian distribution of similar width. The resulting PDF would be the convolution of one of the first distribution with a Gaussian one with arbitrary relative amplitude.}
\label{Chi4PDF}
\end{figure}

The aim of this section is to point to the existence of a regime
where  the statistical properties of the field $\mS_{2}$ are
transformed due to its super-Hubble nonlinear evolution. Such
properties could be transferred in the observed metric perturbation
in case of a small mixing angle.

Let us recall that at super-Hubble scales the evolution equation for
each component of the field is given by,
\begin{equation}
3H\dot\chi_{i}=-\nu^2(\chi_{i}+\chib_{i})\left\vert\chi+\chib\right\vert^2
\end{equation}
for each component of the field and as long as the slow roll
conditions are valid. An equation that obviously  can be solved
explicitly in the classical regime. It leads to,
\begin{equation}
\chi_{i}(t)=\frac{\chi_{i}^{\rm HC}+\chib_{i}}{\left[1+2\nu^2{\left\vert\chi^{\rm HC}+\chib\right\vert^2}\int\frac{\dd t}{3H}\right]^{1/2}}-\chib_{i}
\label{NLtransform}
\end{equation}
where $\chi^{\rm HC}$ is the value of $\chi$ at horizon crossing. It is important to note here that 
large NG will classically develop when $\nu^2 N_e$ is larger then unity whereas the 
self-coupling of $\chi$ at horizon crossing is driven by $\nu^2$. There is thus a domain in parameter space where the nonlinear evolution of the field fluctuations is essentially classical.

Using equation (\ref{NLtransform}) and assuming that $\chi^{\rm HC}$ is Gaussian
distributed, one can compute explicit PDFs of $\chi_{1}$. They are presented on Fig.
\ref{Chi4PDF} for specific values of the $\chib$ parameters. One observes
that for significant values of the coupling constant, the large
excursion values of $\chi_{1}$ are strongly suppressed.

It is then tempting to assume that the resulting field properties
can  derived by the application of such a local transform. That
would indeed lead to interesting phenomenology regarding the local
metric preperties. It is however an approximation the importance of
which is difficult to grasp. Indeed the relation (\ref{NLtransform})
would be valid if the fluctuations had all the same scale and cross
the horizon at the same time. This is not so. For instance when one
observes fluctuation at 10 Mpc scale, its nonlinear evolution is
bound to be affected by modes at much smaller scales, that have
crossed the horizon later, but that nonetheless affect the nonlinear
evolution of the field by changing the local field values. In a
field theory language, this is a UV effect in the radiative
correction. Its importance cannot be neglected since one expects
modes that are up to $e^{60}$ smaller than the observed ones to
cross the horizon during the supre-Hubble evolution of the observed
modes.

The detailed description of these effects is left for a forthcoming
study.  The transform (\ref{NLtransform}) however suggests that the
main effect of such a fully nonlinear evolution of the field is to
squeeze the rare event tails. Such an effect would be visible from
the behavior of the moments of the distributions shown on Fig.
\ref{Chi4PDF}.

%
%
%
%

\section{Conclusions}

We show here that it is possible to build models of multiple field
inflation in a supersymmetric context that can generate a
significant amount of non-Gaussianity. The models we advocate here
are natural extension of D-term inflationary models. As models
producing primordial non-Gaussianities, these types of models fall
in the same class as those described in \cite{2003PhRvD..67l1301B}.
The only difference is that the initial isocurvature fluctuations
are produced by a complex scalar field (not a real) which (mildly)
affects its high-order correlation properties.

Such models are characterized by a mixing angle and an intrinsic
self-coupling parameter. In SUSY context, when one wants to avoid
fine tuned terms in the superpotential,  the potential is naturally
quartic and such models are remarquably constrained. In particular
the shape of the inflationary potential has only a limited number of
parameters. We note that in such a setting the mass of the
isocurvature modes is automatically protected against radiative
corrections (as an application of the results presented in
\cite{2005PhRvD..71f3529B}). In our analysis we have focused on a
regime suggested by the observational constraints regarding cosmic
strings formations. Those constraints impose that the Hubble scale
is essentially constant during the inflationary phase which makes
the transfer of modes, from isocurvature to adiabatic, potentially very efficient.
Finite volume effects also introduce new dynamically determined
parameters that induce effective cubic terms in the potential
\cite{2004PhRvD..70d3533B} the consequence of which we also present.

The phenomenological consequences of such a family of models 
can be obtained analytically.  The
induced high order correlation functions are due to super-Hubble
evolution of the field. This opens the possibility of having weakly
non-Gaussian fields. Two parameter domains can be distinguished for
its phenomenology. If the mixing angle is generic then the current
constraints suggest that the coupling parameter is small enough so
that a tree order calculation of a perturbation theory suffice to
derive the bi- and tri-spectra of the metric field. On the other
hand if the mixing is small enough (0.1 for instance), then the
resulting metric statistical properties might have been shaped by a
nonlinear evolution of the field during its super-Hubble evolution.
In principle, such an evolution can nonetheless be addressed analytically  because, at such scales,
the fields enter a purely classical behavior. We present some
tentative description of the evolution of the high order moments of
the field.

The set of predictions we have obtained has eventually been
confronted to the current constraints and to the expected
constraints that the Planck satellite is expected to provide us for.
It is shown that a joint use of CMB temperature bispectrum and
trispectrum is required to efficiently explore the parameter space
of such models. This, with the curvaton mechanism, is one of the
very few explicit models in which primordial NG can significantly
exceeds those naturally induced by the gravitational dynamics. In this case,
it is fully determined by the inflationary sector of the theory.

\section{Correlators of a test field in de Sitter space}

We review here the formal expression of the connected  bi- and
tri-spectrum of the component of a test scalar complex field in a de
Sitter background. We assume that the potential contains cubic and
quartic terms,
\begin{equation}
V(\chi)=\nu^2\left(\frac{1}{4}\chi_{1}^4+\chi_{1}^3\chib_{1}+\chi_{1}^2\chi_{2}\chib_{2}+\dots
\right)
\end{equation}
as implied by the form (\ref{newpot}). The dots represent terms
that won't affect the (tree order) expressions of the  bi- and
tri-spectrum of $\chi_{1}$.

The perturbative calculations of those correlators can be done  with
the In-In formalism (see
\cite{2003JHEP...05..013M,2005PhRvD..72d3514W}). If $Q$ is the
quantity the vacuum expectation value one wants to compute then it
can be shown that a perturbative expansion is given by
\begin{eqnarray}
&&\mg 0\vert Q(\chi(\eta))\vert 0\md=\int_{\eta_{0}}^{\eta}\dd \eta_{1}
\int_{\eta_{0}}^{{\eta_{1}}}\dd \eta_{2}\dots\nonumber\\
&&\hspace{-1cm}
\mg 0\vert [\dots[[Q(\chi^{(0)}(\eta)), -\ii H^{(I)}(\eta_{1})], -\ii H^{(I)}(\eta_{2})]\dots]\vert 0\md,
\end{eqnarray}
where $H^{(I)}(\eta)$ is the interaction part of the Hamiltonien,
$\eta_{0}$ is an arbitrarily early time when the interaction is
supposed to start playing a role, $\eta$ is the time at which the
expectation value is computed and $\chi^{(0)}(\eta)$ are the field
values at time $\eta$ when they evolve according to the
non-interacting part of the Lagrangian. In the following we use this
formulae for $Q=\chi_{1}(\vk_{1})\dots\chi_{1}(\vk_{3})$,
$Q=\chi_{1}(\vk_{1})\dots\chi_{1}(\vk_{4})$ and for
$H^{(I)}=\int\dd^3\vx \sqrt{-g}\,V(\chi)$ while keeping the
expansion only to tree order.

The expression of those correlators will depend on the time
dependance of the free field. The latter is given by
\begin{equation}
\chi_{i}^{(0)}(\vx)=\frac{1}{a}\int\dd^3\vk \left(f_{k}(\eta) a_{\vk}e^{\ii\vk\vx}+
f_{k}^*(\eta) a^{\dag}_{\vk}e^{-\ii\vk\vx}
\right)
\end{equation}
where,
\begin{equation}
f_{k}=\frac{1}{\sqrt{2\,k}} \left(1+\frac{\ii}{k \eta }\right){e^{\ii k \eta}}
\end{equation}
for a de Sitter background. $\eta$ is here the conformal time, $\eta=-1/(aH)$.
this function permits to define the different time Green function, $G_{k}(\eta,\eta')$,
\begin{equation}
G_{k}(\eta,\eta')=f_{k}(\eta')f_{k}^{*}(\eta).
\end{equation}

\begin{figure*}
\centerline{\epsfig{figure=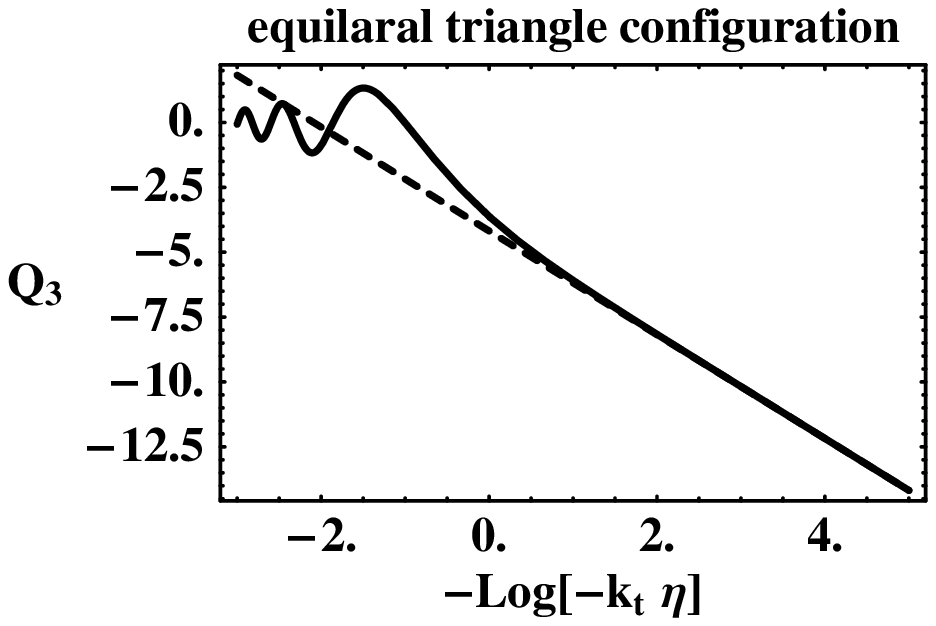,width=8cm}\epsfig{figure=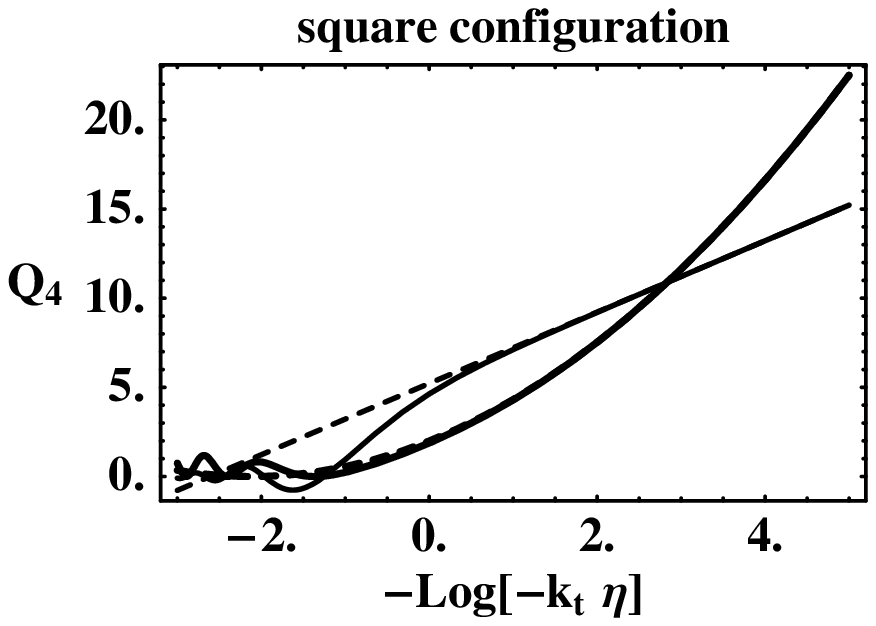,width=8cm}}
\caption{Behavior of the reduced correlateors, $Q_{3}$ and $Q_{4}$ of the field $\chi_{1}$, as a function of $N_{e}\equiv \log(k_{t}\eta)$. The function $Q_{3}$ (left panel) is to be multiplied
by $\nu^2\chib_{1}/H^2$; the function $Q_{4}^{\hbox{star}}$ (thick lines of right panel) is to be multiplied by $-\nu^2/H^2$ and the function $Q_{4}^{\hbox{line}}$  (thin lines) is to be multiplied by $2\nu^4(9\chib_{1}^2+\chib_{2}^2)/H^4$. The dashed lines correspond to the corresponding asymptotic behahiors.}
\label{Qsevol}
\end{figure*}

After some calculations, one can show that the  tree order  term for
the three-point function is given by,
\begin{eqnarray}
P_{3}^{\chi_{1}}(\vk_{1},\vk_{2},\vk_{3})&=&6\ii\,\nu^2\chib_{1}\,\frac{1}{a^3}\nonumber\\
&&\hspace{-2.6cm}\int_{-\infty}^{\eta}\frac{\dd\eta'}{H\eta'}
\left[G_{k_{1}}(\eta,\eta')G_{k_{2}}(\eta,\eta')G_{k_{3}}(\eta,\eta')-
\cc\right].
\end{eqnarray}
It is interesting to define a reduced bispectrum $Q_{3}(\vk_{1},\vk_{2},\vk_{3})$
as
\begin{equation}
Q_{3}(\vk_{1},\vk_{2},\vk_{3})=
\frac{P_{3}^{\chi_{1}}(\vk_{1},\vk_{2},\vk_{3})}{P^{\chi_{1}}(k_{1})P^{\chi_{1}}(k_{2})+\sym}.
\end{equation}
This function is shown on Fig. \ref{Qsevol} as a function of $\eta$,
or more precisely as a function of the number of efolds since (or
before) horizon crossing. The mode configuration correspond to an
equilateral configuration ($k_{1}=k_{2}=k_{3}$) but the asymptotic
behavior of $Q_{3}$ is independent on the configuration (dashed
line) if $N_{e}$ is defined as $N_{e}=-\log(-k_{t}\eta)$ with
$k_{t}=k_{1}+k_{2}+k_{3}$.

For the four point function two types of diagrams are contributing
at tree order. One is due to the quartic part of the potential. In
the following we will denote it as the "star" contribution. The
other is due to the cubic terms of the potential. There are two such
contributions (one due to an exchange of $\chi_{1}$ degree of
freedom and one to an exchange of $\chi_{2}$). Altough a priori
smaller than the previous term, these two contributions are not
necessarily negligeable.

The star contribution is given by \cite{2004PhRvD..69f3520B},
\begin{eqnarray}
P_{\hbox{star}}^{\chi_{1}}(\vk_{1},\vk_{2},\vk_{3},\vk_{4})&=&-6\ii\,\nu^2\,
\frac{1}{a^4}\nonumber\\
&&\hspace{-3.6cm}\int_{-\infty}^{\eta}{\dd\eta'}
\left[G_{k_{1}}(\eta,\eta')G_{k_{2}}(\eta,\eta')G_{k_{3}}(\eta,\eta')G_{k_{4}}(\eta,\eta')-\cc\right].
\end{eqnarray}
The line contribution is finally given by the formal expression,
\begin{widetext}
\begin{eqnarray}
P_{\hbox{line}}^{\chi_{1}}(\vk_{1},\vk_{2},\vk_{3},\vk_{4})&=&-2\nu^4\left(9\chib_{1}^2+\chib_{2}^2\right)\,
\frac{1}{a^4}\nonumber\\
&&\hspace{-3.cm}\int_{-\infty}^{\eta}\frac{\dd\eta_{1}}{H\eta_{1}}
\frac{\dd\eta_{2}}{H\eta_{2}}\left[G_{\vert\vk_{1}+\vk_{2}\vert}(\eta_{1},\eta_{2})+\cc\right]
\left[G_{k_{1}}(\eta,\eta_{1})G_{k_{2}}(\eta,\eta_{1})-\cc\right]\left[G_{k_{3}}(\eta,\eta_{1})G_{k_{4}}(\eta,\eta_{2})-
\cc\right]+\hbox{sym.}.
\end{eqnarray}
\end{widetext}
Similarly to $Q_{3}$ one can define the reduced tri-spectra as\footnote{There is here an ambiguity on how $Q_{4}$ can be defined since the two contributions do not have the same late time super-horizon behavior.},
\begin{eqnarray}
Q_{\hbox{star}}(\vk_{1},\vk_{2},\vk_{3},\vk_{4})&=&\nonumber\\
&&\hspace{-2.5cm}
\frac{P_{\hbox{star}}^{\chi_{1}}(\vk_{1},\vk_{2},\vk_{3},\vk_{4})}{P^{\chi_{1}}(k_{1})P^{\chi_{1}}(k_{2})P^{\chi_{1}}(k_{3})+\sym}.
\end{eqnarray}
and
\begin{eqnarray}
Q_{\hbox{line}}(\vk_{1},\vk_{2},\vk_{3},\vk_{4})&=&\nonumber\\
&&\hspace{-2.5cm}
\frac{P_{\hbox{line}}^{\chi_{1}}(\vk_{1},\vk_{2},\vk_{3},\vk_{4})}{P^{\chi_{1}}(k_{1})P^{\chi_{1}}(k_{3})P^{\chi_{1}}(\vert \vk_{1}+\vk_{2})\vert+\sym}.
\end{eqnarray}
This function is shown on right panel of Fig. \ref{Qsevol} for a
peculiar configuration. Note that the relative importance of the two
depends on the configuration even in the asymptotic limits.

The asymptotic limit of the line part of the tri-spectrum is given by,
\begin{widetext}
\begin{equation}
P_{4}^{\hbox{line}}(\vk_{1},\vk_{2},\vk_{3},\vk_{4})\to
-\frac{\nu^4(9 \chib_{1}^2+\chib_{2}^2)}{72\,H^4\ k^3\, k_{1}^3\,
k_{2}^3 \,k_{3}^3\, k_{4}^3}
\left[q_{4}(\vk_{1},\vk_{2},\eta)q_{4}(\vk_{3},\vk_{4},\eta)+\sym\right]
\end{equation}
with
\begin{eqnarray}
q_4(\vk_1,\vk_4,\eta)&=& 2(k_1+k_2)k^2+(k_1^3+k_2^3)(2-2\gamma)\nonumber\\
&&\hspace{-2cm}+2k_1k_2(k_1^2+k_2^2)+k^3\log[(k_1+k_2+k)/(k_1+k_2-k)]-
(k_1^3+k_2^3)\log[\eta^2(k_1+k_2+k)(k_1+k_2-k)],
\end{eqnarray}
\end{widetext}
where $k=\vert\vk_1+\vk_2\vert$ and $\gamma$ is the Euler number.
Note that the dominant contribution of $q_4$ at super-horizon scale
is due to the last term. It behaves like $-2(k_1^3+k_2^3)N_e$. This
leads to the super-horizon behavior of the four point function of
$\chi_1$ exploited in the text.

\begin{acknowledgments}
FB thanks Galileo Galilei Institute, Florence,
for its hospitality in the course of this work. The authors also thank Jean-Philippe Uzan
for fruitful discussions and comments regarding this work. FB is also partially supported by the the French Programme National de Cosmologie.
\end{acknowledgments}


\end{document}